\documentclass[11pt]{article}
\begin{document}
%\begin{center}
\title{\bf A Model of Perfect Fluid with Spin\\ and Non-Abelian Color Charge}
\author{O. V. Babourova,\thanks{E-mail:baburova@orc.ru}\\
Department of Theoretical Physics, Faculty of Physics,\\
Moscow State University, Leninskie Gory, d.1, st.2, Moscow 119992, Russia,\\
A. S. Vshivtsev,\thanks{Deceased}\\
Academician V. P. Myasnikov,\\
Keldysh Institute of Applied Mathematics, Russian Academy of Sciences,\\
Miusskaya pl. 4, Moscow 125045, Russia,\\
\and B. N. Frolov,\thanks{E-mail:frolovbn@orc.ru}\\
Department of Mathematics, Moscow State Pedagogical University,\\
Krasnoprudnaya 14, Moscow 107140, Russia
}
\date{}
\maketitle
\vskip 0.4cm
\par
    At present, there is a tendency for mutual penetration of the methods of the continuous
medium mechanics and the methods of classical and quantum field theory. On the one hand, the
use of the Yang--Mills formalism in perfect fluid hydrodynamics allows one to construct new
gauge-invariant models \cite{GM}. On the other hand, in field theory, the hydrodynamic approach
(according to which the quantum system of interacting quarks and gluons is approximated classically by
the perfect fluid with spin and non-Abelian color charge) is being successfully developed \cite{HK},\cite{Am}.
In this work, a relativistic variational theory of the perfect spin fluid with color charge in
an external non-Abelian Yang--Mills field will be derived in the Riemann--Cartan space $U_{4}$ with
curvature and torsion, which will take into account the spin polarization chromomagnetic effects.
\par
    As the dynamic variables describing the fluid, we take the null-form fields $\psi$ and their
conjugates $\bar{\psi}$, which are transformed by the representations of the direct product of the
Lorentz group and $SU(3)$ color group. An element of the fluid has a vector of four-dimensional velocity
$\vec{u}=u^{a}\vec{e}_{a}$, whose corresponding three-  and one-forms of velocity may be taken, respectively, as
$u: = \vec{u}\rfloor  \eta  =  u^{a}\eta_{a}$ and $*u = u_{a}\theta^{a}= g(\vec{u},\cdots)$, with
$u \wedge *u = c^{2}\eta$, which is the conventional condition for the squared velocity:
$g(\vec{u},\vec{u}) = - c^{2}$. Here  $\theta^{a}$ are the basis one-forms, $\eta$ is the volume four-form.
$\rfloor$ denotes the inner product, and $*$ denotes the Hodge dualization. We also introduce the three-
and two-form fields \cite{Tr} $\eta_{b}: = \vec{e}_{b}\rfloor \eta = *\theta_{b}$ and
$\eta_{ab}: =  \vec{e}_{b}\rfloor \eta_{a} = *(\theta_{a} \wedge\theta_{b})$, respectively. The basis $\vec{e}_{a}$
is supposed to be nonholonomic orthogonal with $g(\vec{e}_{a},\vec{e}_{b}) =: g_{ab} = \mbox{diag}(1,1,1,-1)$.
\par
    The internal-energy density $\varepsilon$ of the fluid depends on the extensive (additive) thermodynamic
parameters -- the number of particles per unit volume (concentration) $n$ and the entropy per particle $s$,
which are calculated in the intrinsic frame of reference -- and on the quantities describing the internal
degrees of freedom of the fluid particle, namely, the spin tensor $S_{ab} =  \bar{\psi}M_{ab}\psi$ and the color
charge $J_{m} = \bar{\psi} I_{m}\psi$, where $M_{ab}$ and $I_{m}$ are the generators of the corresponding
representations of the Lorentz and $SU(3)$ color groups, respectively. The quantities characterizing the fluid
(including $\psi$ and $\bar{\psi}$) are considered to be averaged over a certain elementary three-volume that
contains what is referred to as a fluid element.
\par
    Furthermore, the fluid element in itself represents a statistical subsystem with a sufficiently large number
of particles to consider it as a quasi-closed system whose properties coincide with the statistical properties of
the fluid in a macroscopic state. For the fluid element, the first law of thermodynamics has the following form:
\begin{equation}
d\varepsilon(n, s, S_{ab}, J_{m}) = \frac{\varepsilon + p}{n} dn + n T ds +
\frac{1}{2} n \omega^{ab}dS_{ab} + n \omega^{m}dJ_{m}\; , \label{eq:term}
\end{equation}
where $p$ is the hydrodynamic pressure in the fluid. In (\ref{eq:term}), the coefficient multiplying $dn$ has the
meaning of the chemical potential, and $\omega^{ab}$ and $\omega^{m}$ describe, respectively, the possible spin
and color-charge exchanges between fluid elements.
\par
    We suppose that the laws of conservation of the number of particles and entropy [which can be expressed by the
equalities $d(n u) = 0$ and $d(n s u) = 0$, where $d$ is the operator of exterior differentiation] are valid.
The first of these equations ensures the continuity of flow lines of the fluid particles. The second equation
expresses the constancy of entropy along the flow lines of the fluid and thus constitutes the second law of
thermodynamics for adiabatic flow of the fluid.
\par
    The fact that the spin tensor meets the Frenkel condition $S_{ab}u^{b}  =  0$ expresses the fundamental
physical property of the tensor, its spacelike character. In terms of the exterior forms, this condition can by
written by $(\vec{e}_{a}\rfloor {\cal S})\wedge u = 0$, where the spin two-form ${\cal S} = (1/2)S_{ab} \theta^{a}
\wedge \theta^{b}$ has been introduced.
\par
    The four-form of Lagrangian density, whose integral determines (the action of the theory, is chosen
by us in the form
\begin{eqnarray}
&&{\cal L}_{fluid} = L_{fluid}\;\eta = -\varepsilon(n,s,\psi,\bar{\psi})\eta +
n\bar{\psi} D\psi \wedge u - \chi n J_{m}{\cal F}^{m}\wedge *{\cal S} +
\nonumber \\
&& + n\lambda_{1}(u\wedge *u - c^{2}\eta) + n u\wedge  d\lambda_{2} +
n \lambda_{3} u\wedge ds + n \zeta^{a}(\vec{e}_{a}\rfloor {\cal S})\wedge u \;.
\label{eq:lag}\end{eqnarray}
\par
    In (\ref{eq:lag}), $n$, $s$, $\psi$, $\bar{\psi}$,  and $u$ are regarded as independent variables describing
the perfect fluid dynamics. In doing so, one should take into account that, in view of (\ref{eq:term}),
the dependence of the internal energy on  $\psi$ and $\bar{\psi}$ is realized only through its dependence on
the spin tensor and color charge: $\varepsilon = \varepsilon(n,s,\bar{\psi}M_{ab}\psi,\bar{\psi}I_{m}\psi )$.
The constraints imposed on the independent variables are taken into account by using the undefined Lagrange multipliers
$\lambda_{1}$, $\lambda_{2}$, $\lambda_{3}$ and $\zeta^{a}$. The second term in (\ref{eq:lag}) is a kinetic term in which
$D$ designates the operation of exterior covariant differentiation with respect to both gauge groups [the Lorentz and
$SU(3)$ color groups]. The term with coupling constant $\chi$ in (\ref{eq:lag}), in which
${\cal F}^{m} = (1/2) F^{m}\!_{ab}\theta^{a}\wedge\theta^{b}$ designates the two-form of strength of
the non-Abelian gauge color field, describes the possible spin polarization chromomagnetic effects.
\par
    On variation of (\ref{eq:lag}) over the undefined Lagrange multipliers, we obtain the corresponding constraints,
and on variation over the dynamic variables describing the perfect fluid, we obtain the variational equations of motion
of the perfect fluid with non-Abelian color charge. Using these equations, together with the values of the undefined
Lagrange multipliers determined by means of these equations, it is possible to show that the four-form of Lagrangian
density (\ref{eq:lag}) is proportional to the hydrodynamic pressure in the fluid ${\cal  L}_{fluid} =  p\eta$.
This indicates that the variational theory of perfect spin fluid with internal color charge transforms into
the variational theory of a conventional perfect fluid as a result of correct passage to the limit.
\par
    We use the equations of motion and the identity $DM_{ab} = 0$ to deduce the law that governs changes
of the spin tensor of the fluid particle:
\begin{equation}
u\wedge  DS_{ab} + \frac{2}{c^{2}}S_{[a}\!^{c} u_{b]}\dot{u}_{c}\eta
= - 2 S_{[a}\!^{f}\Pi^{c}_{b]}\Pi^{d}_{f}(\chi F^{m}\!_{cd}J_{m} +
\omega_{cd})\eta\; . \label{eq:sp}
\end{equation}
Here, the projection tensor is designated by $\Pi^{c}_{b} = \delta^{c}_{b} + c^{-2}u^{c}u_{b}$ and
the Newtonian dot signifies the differentiation that is defined for an arbitrary object $\Phi$ by the equality
$\dot{\Phi}^{a}\!_{b} := *\!(u\wedge D\Phi^{a}\!_{b})$. With the zero right-hand side, the law governing changes
of the spin tensor (\ref{eq:sp}) takes the form of the corresponding Weyssenhoff--Raabe law \cite{RSm} for the
perfect spin fluid.
\par
    To derive the equations for the non-Abelian gauge color field, we should add the four-form of the
color-field Lagrangian density to the four-form of Lagrangian density (\ref{eq:lag}):
\begin{equation}
{\cal L}_{matter} = {\cal L}_{fluid} +{\cal L}_{field} \;, \quad
{\cal L}_{field} = -\frac{\alpha}{2}{\cal F}^{m}\wedge *{\cal F}_{m}
\;. \label{eq:Lm} \end{equation}
Here, $\alpha$ is the coupling constant. Lifting and lowering the indices like $m$ at the gauge field, which are
transformed by an adjoint representation of the $SU(3)$ color group, are effected with the metric tensor
$g_{mn} = - (1/2) c_{m}\!^{p}\!_{q}c_{n}\!^{q}\!_{p}$, where $c_{m}\!^{p}\!_{q}$ are the structure constants
of the $SU(3)$ group.
\par
    The equations for the field are obtained by variation of (\ref{eq:Lm}) over the one-form $A^{m}$ of potential of the
gauge color field:
\begin{equation}
\quad D(\alpha *\!{\cal F}_{m} + \chi n J_{m}*\!{\cal S}) =
n J_{m} u \; . \label{eq:A}
\end{equation}
This equation must be supplemented with the Bianchi identity for a non-Abelian gauge field $ D{\cal F}^{m} = 0$.
\par
    Using the equation of motion for the fluid and identity $ DI_{m} = 0$, we derive the law that governs changes of the
non-Abelian color charge:
\begin{equation}
u\wedge DJ_{m} = - c_{m}\!^{p}\!_{n} J_{p} (\chi {\cal F}^{n}\wedge
*{\cal S} + \omega^{n} \eta )\; . \label{eq:J}
\end{equation}
\par
    The derived classical equations of motion for the non-Abelian color field, which take into account spin and
color degrees of freedom, form a self-consistent set. Equation (\ref{eq:J}) shows that, in the case of a non-Abelian field,
the conservation of color charge along a fluid flow line may be violated because of the spin-chromomagnetic interaction.
It is notable that, in the case of an Abelian field, the conservation of electric charge along a fluid flow line always
takes place, since in that case $c_{m}\!^{p}\!_{n} = 0$.
\par
    According to the general principles of the field theory for the Riemann--Cartan space \cite{Tr}, \cite{Kib}, \cite{Fr},
geometric properties of space-time are defined by the canonical energy-momentum tensor and the spin moment tensor of
the matter. In the exterior form formalism, the role of the latter tensor is played by the three-form of spin moment
$\Delta^{a}\!_{b}$, which is defined as a variational derivative of (\ref{eq:Lm}) with respect to the connection one-form
$\Gamma^{b}\!_{a}$ of the Riemann--Cartan space $U_{4}$. The role of the canonical energy-momentum tensor is played by
the energy-momentum three-form $\Sigma_{a}$, which is defined as a variational derivative of the Lagrangian density
four-form (\ref{eq:Lm}) with respect to the basis one-form $\theta^{a}$. On calculating $\Sigma_{a}$ for the Lagrangian
density (\ref{eq:Lm}) with the use of the equations of motion of the fluid and the expressions for the undefined Lagrangian
multipliers, we arrive at
\begin{eqnarray}
\Sigma_{a} &=& - \frac{\delta{\cal L}_{matter}}{\delta\theta^{a}} =
\Sigma^{fluid}_{a} + \Sigma^{field}_{a}\; , \label{eq:sum}\\
\Sigma^{fluid}_{a} &=& p\eta_{a} + n\left (\pi_{a} +\frac{p}{nc^{2}}
u_{a}\right ) u + \chi n(\vec{e}_{a}\rfloor {\cal F}^{m}J_{m})\wedge *{\cal S}\; ,
\label{eq:sig} \\
\Sigma^{field}_{a} &=& \alpha (F^{m}\!_{ac}F_{m}\!^{bc} - \frac{1}{4}
\delta^{b}_{a}F^{m}\!_{cd}F_{m}\!^{cd})\eta_{b} = \nonumber \\
&& = \frac{\alpha}{2}\left ((\vec{e}_{a}\rfloor {\cal F}^{m}) \wedge *{\cal F}_{m}
- {\cal F}^{m}\wedge (\vec{e}_{a}\rfloor *\!{\cal F}_{m})\right )\; .\nonumber
\end{eqnarray}
In these expressions, the dynamic momentum per particle in a fluid element is introduced:
$$
\pi_{a}\eta = - \frac{1}{nc^{2}}*\!u\wedge \Sigma^{fluid}_{a}\;, \quad
\pi_{a} = \frac{1}{c^{2}}\varepsilon^{*}u_{a} - \frac{1}{c^{2}} S_{a}\!^{c}
\left (\dot{u}_{c} + u^{b}(\chi J_{m}F^{m}\!_{bc} + \omega_{bc})\right ) \; ,
\nonumber
$$
Here  $\varepsilon^{*}$  denotes  the  effective energy per fluid particle, $\varepsilon^{*}\eta = \varepsilon_{o}\eta +
\chi   J_{m}{\cal   F}^{m}\wedge   *{\cal   S}$, where $\varepsilon_{o} = \varepsilon /n$. This expression generalizes the
corresponding result obtained for the case of an electromagnetic field \cite{Cor} to the non-Abelian case. We see that
the role of the fluid pressure $p$ is twofold: first, it prevents the compression of the fluid [the first term in
(\ref{eq:sig})], and, second, it effectively increases the mass of the fluid element [the second term in (\ref{eq:sig})].
\par
    It is well known in the theory of electromagnetic fields that the equations of motion of charged particles
follow from the energy-momentum conservation law of the system together with the field equations, and, in turn,
the energy-momentum conservation law is a consequence of the invariant properties of the system.
In the field theory in $U_{4}$ space, a similar statement is valid, but the laws of conservation have a more
complicated form \cite{Tr}:
\begin{equation}
D\Sigma_{a} = (\vec{e}_{a}\rfloor {\cal T}^{b})\wedge \Sigma_{b} - (\vec{e}_{a}\rfloor
{\cal    R}^{c}\!_{b})\wedge    \Delta^{b}\!_{c}\; , \qquad
D\Delta_{ab}  =  -  \theta_{  [a}\wedge  \Sigma_{b]}\; ,   \label{eq:zak}
\end{equation}
Here, ${\cal T}^{b}$ is the torsion two-form and ${\cal R}^{c}\!_{b}$ is the curvature two-form of $U_{4}$ space.
The first of these equations is a consequence of invariance of the system's Lagrangian under diffeomorphisms of
the energy-momentum, and the second equation is a consequence of local Lorentz invariance and a generalization of the
law of conservation of the spin moment. Both of these equations become identities if and only if the equations of
motion of the system in $U_{4}$ space hold. Thus, these equations are in fact just another representation of the
laws of motion.
\par
    We use the first of equations (\ref{eq:zak}) to derive the equation of motion of the perfect spin fluid with
color charge in the form generalizing the Euler hydrodynamic equation. Substituting the energy-momentum tensor
(\ref{eq:sum}) in (\ref{eq:zak}) and
taking into account the gauge color field equation (\ref{eq:A}), we obtain
\begin{eqnarray}
u\wedge D(\pi_{a} + \frac{p}{nc^{2}}u_{a}) = \frac{1}{n}\eta \nabla_{a}p
- (\vec{e}_{a}\rfloor {\cal T}^{b})\wedge (\pi_{b} + \frac{p}{nc^{2}}u_{b}) u -
\nonumber \\
- (\vec{e}_{a}\rfloor {\cal R}^{bc})\wedge (\frac{1}{2}S_{bc}u) -
(\vec{e}_{a}\rfloor {\cal F}^{m})\wedge (J_{m}u) +
(\nabla_{a}\wedge {\cal F}^{m})
\wedge(\chi J_{m} *\!{\cal S}) , \label{eq:euler}
\end{eqnarray}
where $\nabla_{a}$ designates the covariant differentiation along the direction of the basis field $\vec{e}_{a}$ and
$\nabla_{a}\wedge $ in the last term denotes the covariant differentiation of the tensor-like forms. The third term
in the right-hand side of equation (\ref{eq:euler}) represents the Mathisson force arising from the interaction of the
fluid-particle spin with the space curvature; the second term is characteristic of the $U_{4}$  space and reflects
the interaction between the torsion of the space and the dynamic momentum of the particle. The fourth term in the
right-hand of this equation generalizes the Lorentz force to the case of a non-Abelian gauge field.
\par
    Now let us consider a zero fluid-pressure limiting case. In this case, equation (\ref{eq:euler}) describes the
motion of a single fluid particle that has a spin and color charge. Equations (\ref{eq:euler}), (\ref{eq:sp}) and
(\ref{eq:J}),  together constitute the classical equations of motion of the color panicle, which generalize the Wong
equations \cite{Wo} to the case of the $SU(3)$ color group and allow for the spin of the particles.
\par
    The constructed theory may form the basis of a new model that hydrodynamically describes the quark-gluon plasma
and differs from the Landau model in that it allows for the structure of quantum-chromodynamics vacuum and color
charge of quarks in an explicit form \cite{VP}. Moreover, on the basis of the theory of perfect spin-dilaton fluid
developed in \cite{BFr}, this model can be generalized to the case of a fluid possessing the dilaton charge, apart
from the spin and the color charge, and thus provide the foundation for studies of the non-perturbed gluon condensate
up to the phase transition from hadron matter to the quark-gluon plasma.

\end{document}